\documentclass[11pt]{article}
\usepackage{colacl06}
\usepackage{times}
\usepackage{latexsym}
\usepackage{tipa}
\usepackage{psfig}
\usepackage[lined,algonl,boxed]{algorithm2e}
\title{Analysis and Synthesis of the Distribution of Consonants over Languages: \\
A Complex Network Approach}

\author{Monojit Choudhury \and Animesh Mukherjee \and Anupam Basu \and Niloy Ganguly \\
Department of Computer Science and Engineering,\\
Indian Institute of Technology Kharagpur\\ 
 \tt \{monojit,animeshm,anupam,niloy\}@cse.iitkgp.ernet.in
}
  
\date{}

\begin{document}
\maketitle
\begin{abstract}
Cross-linguistic similarities are reflected by the speech sound systems of languages all over the world. In this work 
we try to model such similarities observed in the consonant inventories, through a complex bipartite network. We 
present a systematic study of some of the appealing features of these inventories with the help of the bipartite 
network. An important observation is that the occurrence of consonants follows a two regime power law distribution. We  
find that the consonant inventory size distribution together with the principle of preferential attachment are the 
main reasons behind the emergence of such a two regime behavior. In order to further support our explanation we 
present a synthesis model for this network based on the general theory of preferential attachment.         
\end{abstract}

\section{Introduction}\label{intro}

Sound systems of the world's languages show remarkable regularities. Any arbitrary set of consonants and vowels does 
not make up the sound system of a particular language. Several lines of research suggest that cross-linguistic 
similarities get reflected in the consonant and vowel inventories of the languages all over the 
world~\cite{Greenberg:66,Pinker:94,Ladefoged:96}. Previously it has been argued that these similarities are the 
results of certain general principles like {\em maximal perceptual contrast}~\cite{Lindblom:88}, {\em feature 
economy}~\cite{Martinet:68,Boersma:98,Clements:04} and {\em robustness}~\cite{Jakobson:56,Chomsky:68}. Maximal 
perceptual contrast between the phonemes of a language is desirable for proper perception in a noisy environment. In 
fact the organization of the vowel inventories across languages has  been satisfactorily explained in terms of the 
single principle of maximal perceptual contrast~\cite{Jakobson:41,Wang:68}.

There have been several attempts to reason the observed patterns in consonant inventories since 
1930s~\cite{Trub:39,Lindblom:88,Boersma:98,Flemming:02,Clements:04}, but unlike the case of vowels, the structure of 
consonant inventories lacks a complete and holistic explanation~\cite{Boer:00}. Most of the works are confined to 
certain individual principles~\cite{Abry:03,Hinskens:03} rather than formulating a general theory describing the 
structural patterns and/or their stability. Thus, the structure of the consonant inventories continues to be a {\em 
complex} jigsaw puzzle, though the parts and pieces are known.

In this work we attempt to represent the cross-linguistic similarities that exist in the consonant inventories of the 
world's languages through a {\em bipartite network} named {\bf PlaNet} (the {\bf P}honeme {\bf L}anguage {\bf 
N}etwork). PlaNet has two different sets of nodes, one labeled by the languages while the other labeled by the 
consonants. Edges run between these two sets depending on whether or not a particular consonant occurs in a particular 
language. This representation is motivated by similar modeling of certain complex phenomena observed in nature and 
society, such as,
\begin{itemize}
\item Movie-actor network, where movies and actors constitute the two partitions and an edge between them signifies 
that a particular actor acted in a particular movie~\cite{Ramasco:04}.
\item Article-author network, where the edges denote which person has authored which articles~\cite{Newman:01}.
\item Metabolic network of organisms, where the corresponding partitions are chemical compounds and metabolic 
reactions. Edges run between partitions depending on whether a particular compound is a substrate or result of a 
reaction~\cite{Jeong:00}.  
\end{itemize}

Modeling of complex systems as networks has proved to be a comprehensive and emerging way of capturing the underlying 
generating mechanism of such systems (for a review on complex networks and their generation 
see~\cite{Albert:02,Newman:03}). There have been some attempts as well to model the intricacies of human languages 
through complex networks. Word networks based on synonymy~\cite{Yook:01}, co-occurrence~\cite{Cancho:01}, and phonemic 
edit-distance~\cite{Vitevitch:05} are examples of such attempts. The present work also uses the concept of complex 
networks to develop a platform for a holistic analysis as well as synthesis of the distribution of the consonants 
across the languages.  
 
In the current work, with the help of PlaNet we provide a systematic study of certain interesting features of the 
consonant inventories. An important property that we observe is the two regime power law degree 
distribution\footnote{Two regime power law distributions have also been observed in syntactic networks of 
words~\cite{Cancho:01}, network of mathematics collaborators~\cite{Grossman:95}, and language diversity over 
countries~\cite{Gomes:99}.} of the nodes labeled by the consonants. We try to explain this property in the light of 
the size of the consonant inventories coupled with the principle of {\em preferential attachment}~\cite{Albert:99}. 
Next we present a simplified mathematical model explaining the emergence of the two regimes. In order to support our 
analytical explanations, we also provide a synthesis model for PlaNet. 

The rest of the paper is organized into five sections. In section~\ref{def} we formally define PlaNet, outline its 
construction procedure and present some studies on its degree distribution. We dedicate section~\ref{exp_an} to state 
and explain the inferences that can be drawn from the degree distribution studies of PlaNet. In section~\ref{theory} 
we provide a simplified theoretical explanation of the analytical results obtained. In section~\ref{pam} we present a 
synthesis model for PlaNet to hold up the inferences that we draw in section~\ref{exp_an}. Finally we conclude in 
section~\ref{conc} by summarizing our contributions, pointing out some of the implications of the current work and 
indicating the possible future directions.

\section{PlaNet: The Phoneme-Language Network}\label{def}

\begin{figure}
\begin{center}
\psfig{file=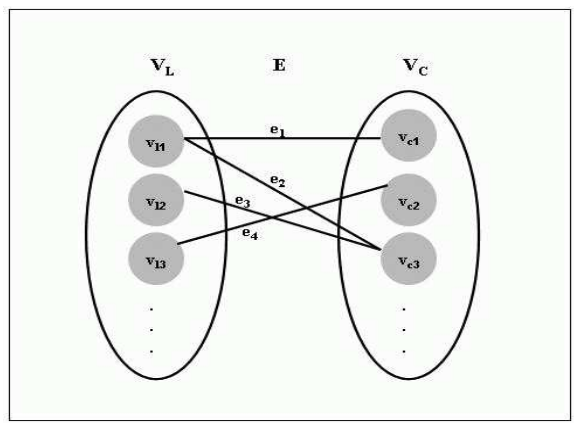,width=3in}
\caption{Illustration of the nodes and edges of PlaNet}
\label{planet}
\end{center}
\end{figure}

We define the network of consonants and languages, PlaNet, as a {\em bipartite graph} represented as G = 
$\langle${V$_L$, V$_C$, E}$\rangle$ where V$_L$ is the set of {\em nodes} labeled by the languages and V$_C$ is the 
set of nodes labeled by the consonants. E is the set of edges that run between V$_L$ and V$_C$. There is an {\em edge} 
$e$ $\in$ E between two nodes $v_l$ $\in$ V$_L$ and $v_c$ $\in$ V$_C$ if and only if the consonant {\em c} occurs in 
the language {\em l}. Figure~\ref{planet} illustrates the nodes and edges of PlaNet.   

\subsection{Construction of PlaNet}

Many typological studies~\cite{Lindblom:88,Ladefoged:96,Hinskens:03}  of segmental inventories have been carried out 
in past on the UCLA Phonological Segment Inventory Database (UPSID)~\cite{Maddieson:84}. UPSID initially had 317 
languages and was later extended to include 451 languages covering all the major language families of the world. In 
this work we have used the older version of UPSID comprising of 317 languages and 541 consonants (henceforth 
UPSID$_{317}$), for constructing PlaNet. Consequently, there are 317 elements (nodes) in the set V$_L$ and 541 
elements (nodes) in the set V$_C$. The number of elements (edges) in the set E as computed from PlaNet is 7022. At 
this point it is important to mention that in order to avoid any confusion in the construction of PlaNet we have 
appropriately filtered out the anomalous and the ambiguous segments~\cite{Maddieson:84} from it. We have completely 
ignored the anomalous segments from the data set (since the existence of such segments is doubtful), and included the 
ambiguous ones as separate segments because there are no descriptive sources explaining how such ambiguities might be 
resolved. A similar approach has also been described in Pericliev and Vald{\'e}s-P{\'e}rez~\shortcite{Peri:02}. 

\subsection{Degree Distribution of PlaNet}

\begin{figure}
\begin{center}
\psfig{file=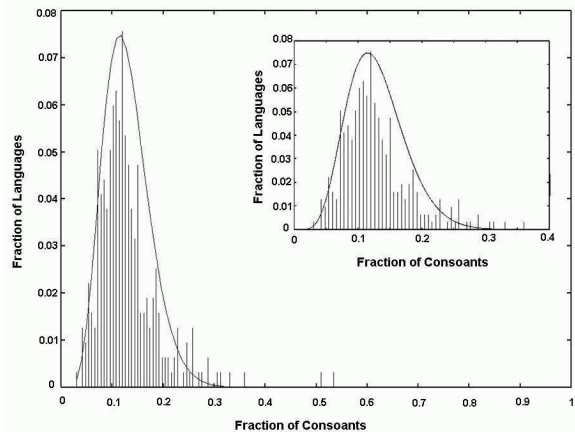,width=3in}
\caption{Degree distribution of PlaNet for the set V$_L$. The figure in the inner box is a magnified version of a 
portion of the original figure.}
\label{dd_lang}
\end{center}
\end{figure}

The {\em degree} of a node $u$, denoted by $k_u$ is defined as the number of edges connected to $u$. The term {\em 
degree distribution} is used to denote the way degrees ($k_u$) are distributed over the nodes ($u$). The degree 
distribution studies find a lot of importance in understanding the complex topology of any large network, which is 
very difficult to visualize otherwise. Since PlaNet is bipartite in nature it has two degree distribution curves one 
corresponding to the nodes in the set V$_L$  and the other corresponding to the nodes in the set V$_C$. 
\paragraph{\bf {Degree distribution of the nodes in V$_L$: }}
Figure~\ref{dd_lang} shows the degree distribution of the nodes in V$_L$ where the x-axis denotes the degree of each 
node expressed as a fraction of the maximum degree and the y-axis denotes the number of nodes having a given degree  
expressed as a fraction of the total number of nodes in V$_L$ . 

\begin{figure}
\begin{center}
\psfig{file=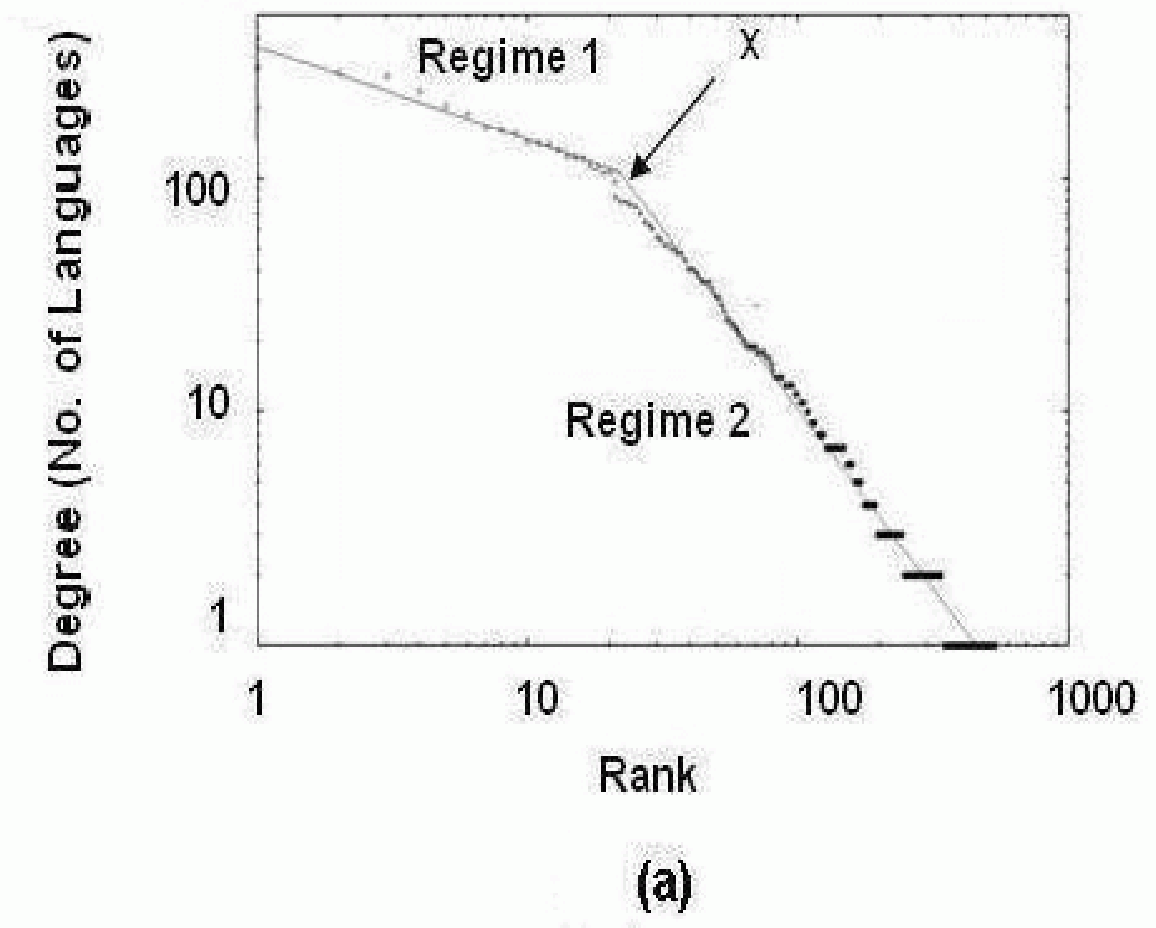,width=2.5in}
\psfig{file=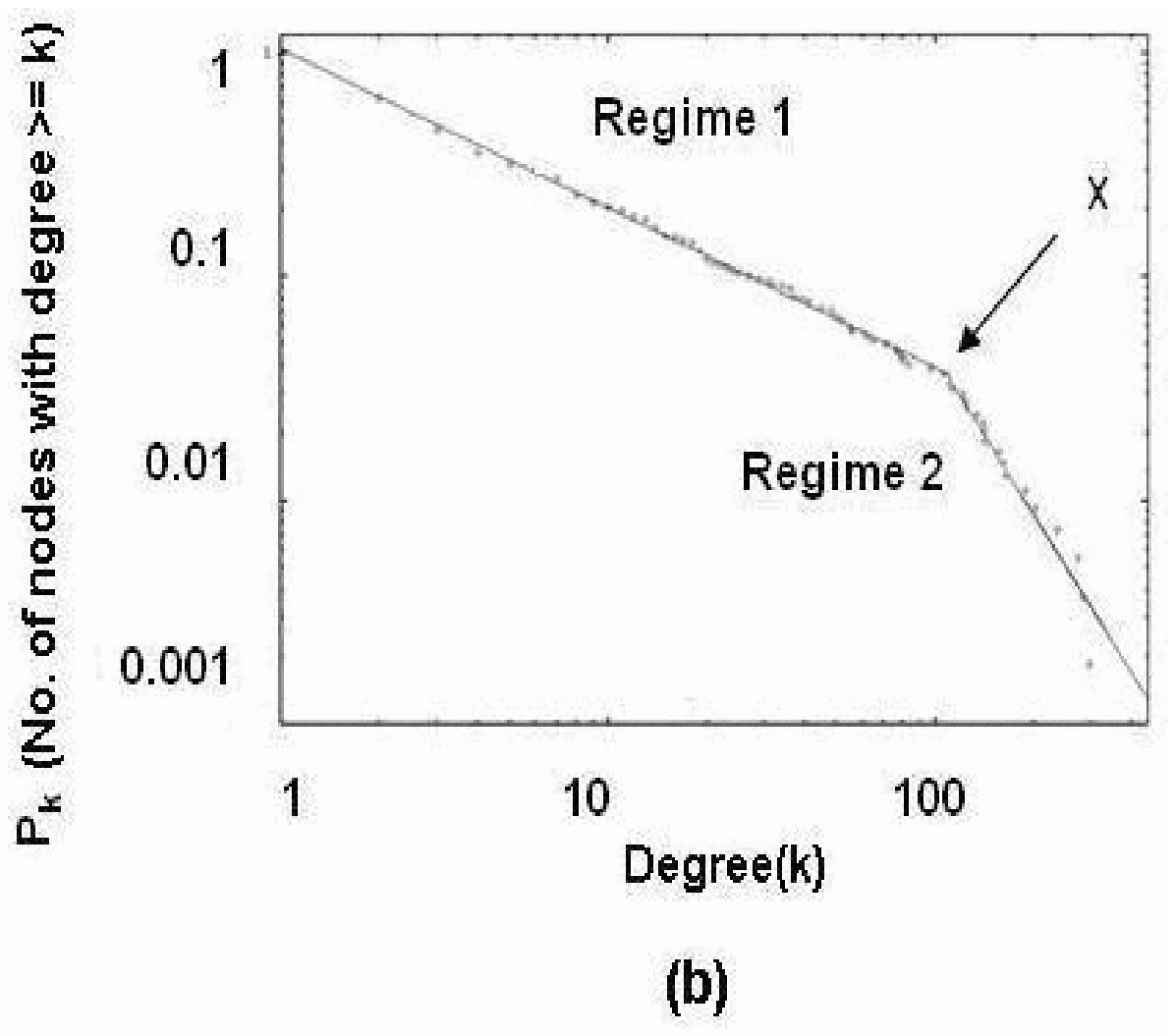,width=2.3in}
\caption{Degree distribution of PlaNet for the set V$_C$ in a log-log scale}
\label{dd_cons}
\end{center}
\end{figure}

It is evident from Figure~\ref{dd_lang} that the number of consonants appearing in different languages follow a 
$\beta$-distribution \footnote{A random variable is said to have a $\beta$-distribution with parameters $\alpha$ $>$ 0 
and $\beta$ $>$ 0 if and only if its probability mass function is given by
\begin{displaymath}
f(x) = \frac{\Gamma({\alpha} + {\beta})}{{\Gamma} ({\alpha}) {\Gamma} ({\beta})}x^{\alpha-1}(1-x)^{\beta-1} 
\end{displaymath} for 0 $<$ x $<$ 1 and $f(x)$ = 0 otherwise. $\Gamma$($\cdot$) is the Euler's gamma function.} 
(see~\cite{Bulmer:79} for reference). The figure shows an asymmetric right skewed distribution with the values of 
$\alpha$ and $\beta$ equal to 7.06 and 47.64 (obtained using maximum likelihood estimation method) respectively. The 
asymmetry points to the fact that languages usually tend to have smaller consonant inventory size, the best value 
being somewhere between 10 and 30. The distribution peaks roughly at 21 indicating that majority of the languages in 
UPSID$_{317}$ have a consonant inventory size of around 21 consonants. 
\paragraph{\bf {Degree distribution of the nodes in V$_C$: }}
Figure~\ref{dd_cons} illustrates two different types of degree distribution plots for the nodes in V$_C$; 
Figure~\ref{dd_cons}(a) corresponding to the rank, i.e., the sorted order of degrees, (x-axis) versus degree (y-axis) 
and Figure~\ref{dd_cons}(b) corresponding to the degree ($k$) (x-axis) versus $P_k$ (y-axis) where $P_k$ is the 
fraction of nodes having degree greater than or equal to $k$. 

Figure~\ref{dd_cons} clearly shows that both the curves have two distinct regimes and the distribution is scale-free.  
Regime 1 in Figure~\ref{dd_cons}(a) consists of 21 consonants which have a very high frequency (i.e., the degree $k$) 
of occurrence. Regime 2 of Figure~\ref{dd_cons}(b) also correspond to these 21 consonants. On the other hand Regime 2 
of Figure~\ref{dd_cons}(a) as well as Regime 1 of Figure~\ref{dd_cons}(b) comprises of the rest of the consonants. The 
point marked as {\bf x} in both the figures indicates the breakpoint. Each of the regime in both 
Figure~\ref{dd_cons}(a) and (b) exhibit a power law of the form 
\begin{displaymath}
y = Ax^{-\alpha}
\end{displaymath}
In Figure~\ref{dd_cons}(a) $y$ represents the degree $k$ of a node corresponding to its rank $x$ whereas in Figure 
~\ref{dd_cons}(b) $y$ corresponds to $P_k$ and $x$, the degree $k$. The values of the parameters A and $\alpha$, for 
Regime 1 and Regime 2 in both the figures, as computed by the least square error method, are shown in 
Table~\ref{param}.

\protect\begin{table*}\centering
\begin{tabular}{|l|l|l|}
\cline{1-3}
\vbox to1.97ex{\vspace{1pt}\vfil\hbox to9.40ex{\hfil Regime \hfil}} & 
\vbox to1.97ex{\vspace{1pt}\vfil\hbox to25.80ex{\hfil Figure~\ref{dd_cons}(a)\hfil}} & 
\vbox to1.97ex{\vspace{1pt}\vfil\hbox to24.60ex{\hfil Figure~\ref{dd_cons}(b)\hfil}} \\

\cline{1-3}
\vbox to1.70ex{\vspace{1pt}\vfil\hbox to9.40ex{\hfil Regime 1\hfil}} & 
\vbox to1.70ex{\vspace{1pt}\vfil\hbox to25.80ex{ A = 368.70 \hfil  $\alpha$ = 0.4\hfil}} & 
\vbox to1.70ex{\vspace{1pt}\vfil\hbox to24.60ex{ A = 1.040 \hfil $\alpha$ = 0.71\hfil}} \\

\cline{1-3}
\vbox to1.70ex{\vspace{1pt}\vfil\hbox to9.40ex{\hfil Regime 2\hfil}} & 
\vbox to1.70ex{\vspace{1pt}\vfil\hbox to25.80ex{ A = 12456.5 \hfil $\alpha$ = 1.54\hfil}} & 
\vbox to1.70ex{\vspace{1pt}\vfil\hbox to24.60ex{ A = 2326.2 \hfil $\alpha$ = 2.36\hfil}} \\

\cline{1-3}
\end{tabular}
\caption{The values of the parameters A and $\alpha$}
\label{param}
\end{table*}

It becomes necessary to mention here that such power law distributions, known variously as Zipf's law~\cite{Zipf:49}, 
are also observed in an extraordinarily diverse range of phenomena including the frequency of the use of words in 
human language~\cite{Zipf:49}, the number of papers scientists write~\cite{Lotka:26}, the number of hits on web 
pages~\cite{Adamic:00} and so on. Thus our inferences, detailed out in the next section, mainly centers around this 
power law behavior.    

\section{Inferences Drawn from the Analysis of PlaNet}\label{exp_an}

In most of the networked systems like the society, the Internet, the World Wide Web, and many others, power law degree 
distribution emerges for the phenomenon of preferential attachment, i.e., when ``the rich get richer"~\cite{Simon:55}. 
With reference to PlaNet this preferential attachment can be interpreted as the tendency of a language to choose a 
consonant that has been already chosen by a large number of other languages. We posit that it is this preferential 
property of languages that results in the power law degree distributions observed in Figure~\ref{dd_cons}(a) and (b).    

\begin{figure}
\begin{center}
\framebox{
\psfig{file=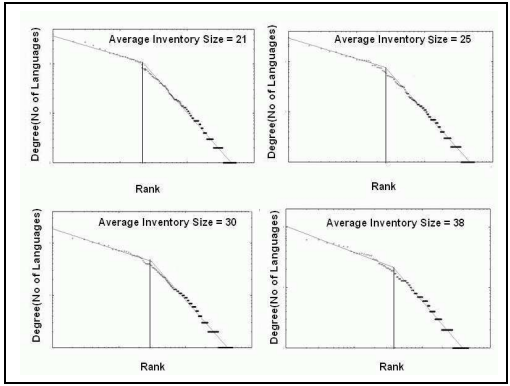,width=2.5in}}
\caption{Degree distributions at different average consonant inventory sizes}
\label{shift}
\end{center}
\end{figure}

Nevertheless there is one question that still remains unanswered. Whereas the power law distribution is well 
understood, the reason for the two distinct regimes (with a sharp break) still remains unexplored. We hypothesize 
that,\\        
{\bf Hypothesis} {\em The typical distribution of the consonant inventory size over languages coupled with the 
principle of preferential attachment enforces the two distinct regimes to appear in the power law curves.}\\     
As the average consonant inventory size in UPSID$_{317}$ is 21, so following the principle of preferential attachment, 
on an average, the first 21 most frequent consonants are much more preferred than the rest. Consequently, the nature 
of the frequency distribution for the highly frequent consonants is different from the less frequent ones, and hence 
there is a transition from Regime 1 to Regime 2 in the Figure~\ref{dd_cons}(a) and (b).
\paragraph{{\bf Support Experiment: }} In order to establish that the consonant inventory size plays an important role 
in giving rise to the two regimes discussed above we present a support experiment in which we try to observe whether 
the breakpoint {\bf x} shifts as we shift the average consonant inventory size.\\
{\bf Experiment:} In order to shift the average consonant inventory size from 21 to 25, 30 and 38 we neglected the 
contribution of the languages with consonant inventory size less than $n$ where $n$ is 15, 20 and 25 respectively and 
subsequently recorded the degree distributions obtained each time. We did not carry out our experiments for average 
consonant inventory size more than 38 because the number of such languages are very rare in UPSID$_{317}$.\\
{\bf Observations:} Figure~\ref{shift} shows the effect of this shifting of the average consonant inventory size on 
the rank versus degree distribution curves. Table~\ref{transition} presents the results observed from these curves 
with the left column indicating the average inventory size and the right column the breakpoint {\bf x}. The table 
clearly indicates that the transition occurs at values corresponding to the average consonant inventory size in each 
of the three cases.\\
{\bf Inferences:} It is quite evident from our observations that the breakpoint {\bf x} has a strong correlation with 
the average consonant inventory size, which therefore plays a key role in the emergence of the two regime degree 
distribution curves. 

In the next section we provide a simplistic mathematical model for explaining the two regime power law with a 
breakpoint corresponding to the average consonant inventory size.   

\protect            

\protect\begin{table}\centering
\begin{tabular}{|l|l|}
\cline{1-2}
\vbox to1.88ex{\vspace{1pt}\vfil\hbox to22.60ex{\hfil Avg. consonant inv. size\hfil}\vfil} & 
\vbox to1.88ex{\vspace{1pt}\vfil\hbox to12.80ex{\hfil Transition\hfil}\vfil} \\

\cline{1-2}
\vbox to1.88ex{\vspace{1pt}\vfil\hbox to22.60ex{\hfil 25\hfil}\vfil} & 
\vbox to1.88ex{\vspace{1pt}\vfil\hbox to12.80ex{\hfil 25\hfil}\vfil} \\

\cline{1-2}
\vbox to1.88ex{\vspace{1pt}\vfil\hbox to22.60ex{\hfil 30\hfil}\vfil} & 
\vbox to1.88ex{\vspace{1pt}\vfil\hbox to12.80ex{\hfil 30\hfil}\vfil} \\

\cline{1-2}
\vbox to1.88ex{\vspace{1pt}\vfil\hbox to22.60ex{\hfil 38\hfil}\vfil} & 
\vbox to1.88ex{\vspace{1pt}\vfil\hbox to12.80ex{\hfil 37\hfil}\vfil} \\

\cline{1-2}
\end{tabular}
\caption{The transition points for different average consonant inventory size}
\label{transition}
\end{table}

\section{Theoretical Explanation for the Two Regimes}\label{theory} 

\begin{figure}
\begin{center}
\psfig{file=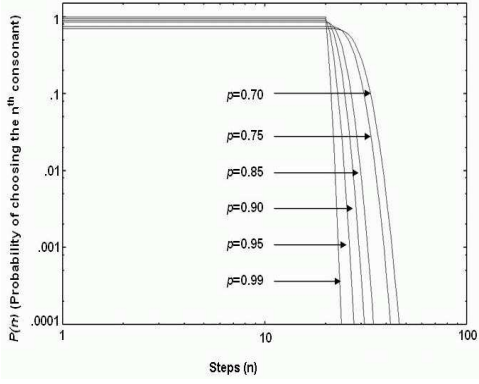,width=2.5in}
\caption{Plot of the function $P(n)$ in log-log scale}
\label{theoretical}
\end{center}
\end{figure}

Let us assume that the inventory of all the languages comprises of 21 consonants. We further assume that the 
consonants are arranged in their hierarchy of preference. A language traverses the hierarchy of consonants and at 
every step decides with a probability $p$ to choose the current consonant. It stops as soon as it has chosen all the 
21 consonants. Since languages must traverse through the first 21 consonants regardless of whether the previous 
consonants are chosen or not, the probability of choosing any one of these 21 consonants must be $p$. But the case is 
different for the 22$^{nd}$ consonant, which is chosen by a language if it has previously chosen zero, one, two, or at 
most 20, but not all of the first 21 consonants. Therefore, the probability of the 22$^{nd}$ consonant being chosen 
is,
\begin{displaymath}
P(22) = p\sum_{i=0}^{20}\left(\begin{array}{c}21\\ i\end{array}\right)p^i(1-p)^{21-i}\\
\end{displaymath}   
where
\begin{displaymath}
 \left(\begin{array}{c}21\\ i\end{array}\right)p^i(1-p)^{21-i}
\end{displaymath}  
denotes the probability of choosing $i$ consonants from the first 21. In general the probability of choosing the 
n+1$^{th}$ consonant from the hierarchy is given by, 
\begin{displaymath}
P(n+1) = p\sum_{i=0}^{20}\left(\begin{array}{c}n\\ i\end{array}\right)p^i(1-p)^{n-i}
\end{displaymath}  
Figure~\ref{theoretical} shows the plot of the function $P(n)$ for various values of $p$ which are 0.99, 0.95, 0.9, 
0.85, 0.75 and 0.7 respectively in log-log scale. All the curves, for different values of $p$, have a nature similar 
to that of the degree distribution plot we obtained for PlaNet. This is indicative of the fact that languages choose 
consonants from the hierarchy with a probability function comparable to $P(n)$.

Owing to the simplified assumption that all the languages have only 21 consonants, the first regime is a straight 
line; however we believe a more rigorous mathematical model can be built taking into consideration the 
$\beta$-distribution rather than just the mean value of the inventory size that can explain the negative slope of the 
first regime. We look forward to do the same as a part of our future work. Rather, here we try to investigate the 
effect of the exact distribution of the language inventory size on the nature of the degree distribution of the 
consonants through a synthetic approach based on the principle of preferential attachment, which is described in the 
subsequent section. 
       
\section{The Synthesis Model based on Preferential Attachment}\label{pam}

Albert and Barab{\'a}si~\shortcite{Albert:99} observed that a common property of many large networks is that the 
vertex connectivities follow a scale-free power law distribution. They remarked that two generic mechanisms can be 
considered to be the cause of this observation: (i) networks expand continuously by the addition of new vertices, and 
(ii) new vertices attach preferentially to sites (vertices) that are already well connected. They found that a model 
based on these two ingredients reproduces the observed stationary scale-free distributions, which in turn indicates 
that the development of large networks is governed by robust self-organizing phenomena that go beyond the particulars 
of the individual systems.

\begin{figure}[!h]
\begin{center}
\framebox{
\psfig{file=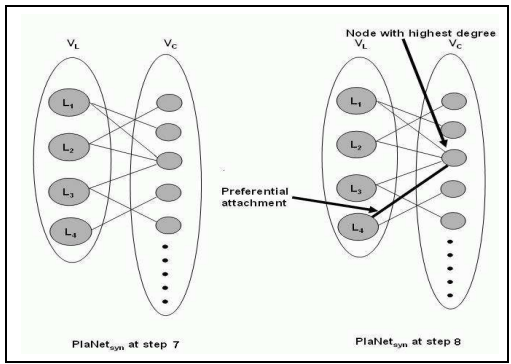,width=2.5in}}
\caption{A partial step of the synthesis process. When the language L$_4$ has to connect itself with one of the nodes 
in the set V$_C$ it does so with the one having the highest degree (=3) rather than with others in order to achieve 
preferential attachment which is the working principle of our algorithm}
\label{planetsyn}
\end{center}
\end{figure}

Inspired by their work and the empirical as well as the mathematical analysis presented above, we propose a 
preferential attachment model for synthesizing PlaNet (PlaNet$_{syn}$ henceforth) in which the degree distribution of 
the nodes in V$_L$ is known. Hence V$_L$=\{L$_1$, L$_2$, . . ., L$_{317}$\} have degrees (consonant inventory size) 
\{$k_1$, $k_2$, . . ., $k_{317}$\} respectively. We assume that the nodes in the set V$_C$ are {\em unlabeled}. At 
each time step, a node L$_j$ ($j$ = 1 to 317) from V$_L$ tries to attach itself with a new node $i \in$ V$_C$ to which 
it is not already connected.  The probability $Pr(i)$ with which the node L$_j$ gets attached to $i$ depends on the 
current degree of $i$ and is given by 
\begin{displaymath}
 Pr(i) = \frac{k_{i} + \epsilon}{\sum_{{i^{'}} \in V_{j}} (k_{i^{'}} + \epsilon)} 
\end{displaymath} where $k_{i}$ is the current degree of the node $i$, V$_{j}$ is the set of nodes in V$_C$ to which 
L$_j$ is not already connected and $\epsilon$ is the smoothing parameter which is used to reduce bias and favor at 
least a few attachments with nodes in V$_{j}$ that do not have a high $Pr(i)$. The above process is repeated until all 
L$_j \in$ V$_L$ get connected to exactly $k_j$ nodes in V$_C$. The entire idea is summarized in Algorithm~\ref{syn}. 
Figure~\ref{planetsyn} shows a partial step of the synthesis process  illustrated in Algorithm~\ref{syn}. 

\protect\begin{algorithm}[t]
\BlankLine
  \Repeat{all languages complete their inventory quota}{
  \For{j = 1 to 317}{
 \If{there is a node L$_j$ $\in$ V$_L$ with at least one or more consonants to be chosen from V$_C$}{\
 \BlankLine 
   Compute $V_{j}$ = $V_C$-$V(L_j)$, where $V(L_j)$ is the set of nodes in $V_C$ to which $L_j$ is already connected;
    }
 \BlankLine 
   \For{each node i $\in$ V$_{j}$}{
    \begin{displaymath}
    Pr(i) = \frac{k_{i} + \epsilon}{\sum_{{i^{'}} \in V_{j}} (k_{i^{'}} + \epsilon)} 
	\end{displaymath} where $k_{i}$ is the current degree of the node $i$ and $\epsilon$ is the model parameter. 
$Pr(i)$ is the probability of connecting $L_j$ to $i$.
    }
    \BlankLine
	Connect $L_j$ to a node $i \in V_{j}$ following the distribution $Pr(i)$;
 } 
} 
\label{syn}
\caption{Algorithm for synthesis of PlaNet based on preferential attachment}    
\end{algorithm}

\begin{figure}[!t]
\begin{center}
\psfig{file=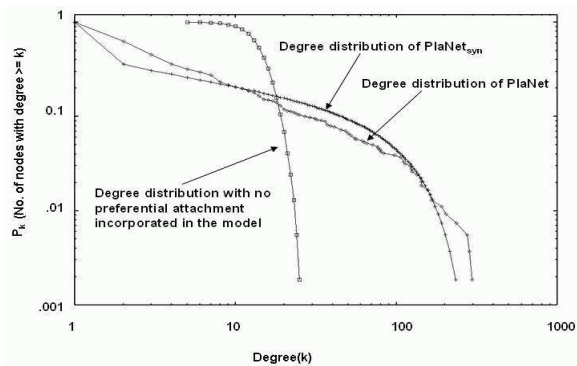,width=3in}
\caption{Degree distribution of the nodes in V$_C$ for both PlaNet$_{syn}$, PlaNet, and when the model incorporates no 
preferential attachment; for PlaNet$_{syn}$, $\epsilon$ = 0.0701 and the results are averaged over 100 simulation 
runs}
\label{deg_synth}
\end{center}
\end{figure}

\paragraph{{\bf Simulation Results: }}Simulations reveal that for PlaNet$_{syn}$ the degree distribution of the nodes 
belonging to V$_C$ fit well with the analytical results we obtained earlier in section~\ref{def}. Good fits emerge for 
the range 0.06 $\leq \epsilon \leq$ 0.08 with the best being at $\epsilon$ = 0.0701. Figure~\ref{deg_synth} shows the 
degree $k$ versus $P_k$ plots for $\epsilon$ = 0.0701 averaged over 100 simulation runs.  

The mean error\footnote{Mean error is defined as the average difference between the ordinate pairs where the abscissas 
are equal.} between the degree distribution plots of PlaNet and PlaNet$_{syn}$ is 0.03 which intuitively signifies 
that on an average the variation in the two curves is 3$\%$. On the contrary, if there were no preferential attachment 
incorporated in the model (i.e., all connections were equiprobable) then the mean error would have been 0.35 (35$\%$ 
variation on an average).   

\section{Conclusions, Discussion and Future Work}\label{conc}

In this paper, we have analyzed and synthesized the consonant inventories of the world's languages in terms of a 
complex network. We dedicated the preceding sections essentially to,
\begin{itemize}
\item Represent the consonant inventories through a bipartite network called PlaNet,
\item Provide a systematic study of certain important properties of the consonant inventories with the help of PlaNet,
\item Propose analytical explanations for the two regime power law curves (obtained from PlaNet) on the basis of the 
distribution of the consonant inventory size over languages together with the principle of preferential attachment,    
\item Provide a simplified mathematical model to support our analytical explanations, and
\item Develop a synthesis model for PlaNet based on preferential attachment where the consonant inventory size 
distribution is known {\em a priori}.
\end{itemize}

We believe that the general explanation provided here for the two regime power law is a fundamental result, and can 
have a far reaching impact, because two regime behavior is observed in many other networked systems.

Until now we have been mainly dealing with the computational aspects of the distribution of consonants over the 
languages rather than exploring the real world dynamics that gives rise to such a distribution. An issue that draws 
immediate attention is that how preferential attachment, which is a general phenomenon associated with network 
evolution, can play a prime role in shaping the consonant inventories of the world's languages. The answer perhaps is 
hidden in the fact that language is an evolving system and its present structure is determined by its past 
evolutionary history. Indeed an explanation based on this evolutionary model, with an initial disparity in the 
distribution of consonants over languages, can be intuitively verified as follows -- let there be a language community 
of N speakers communicating among themselves by means of only two consonants say /$k$/ and /$g$/. If we assume that 
every speaker has $l$ descendants and language inventories are transmitted with high fidelity, then after $i$ 
generations it is expected that the community will consist of $ml^i$ /$k$/ speakers and $nl^i$ /$g$/ speakers. Now if 
$m > n$ and $l >$ 1, then for sufficiently large $i$, $ml^i \gg nl^i$. Stated differently, the /$k$/ speakers by far 
outnumbers the /$g$/ speakers even if initially the number of /$k$/ speakers is only slightly higher than that of the 
/$g$/ speakers. This phenomenon is similar to that of preferential attachment where language communities get attached 
to, i.e., select, consonants that are already highly preferred. Nevertheless, it remains to be seen where from such an 
initial disparity in the distribution of the consonants over languages might have originated. 

In this paper, we mainly dealt with the occurrence principles of the consonants in the inventories of the world's 
languages. The work can be further extended to identify the co-occurrence likelihood of the consonants in the language 
inventories and subsequently identify the groups or communities within them. Information about such communities can 
then help in providing an improved insight about the organizing principles of the consonant inventories.


\begin{thebibliography}{}

\bibitem[\protect\citename{Abry}2003]{Abry:03}
C. Abry.
\newblock 2003.
\newblock [b]-[d]-[g] as a universal triangle as
acoustically optimal as [i]-[a]-[u]. {\em 15th Int. Congr. Phonetics
Sciences ICPhS}, 727--730.

\bibitem[\protect\citename{Adamic and Huberman}2000]{Adamic:00}
L. A. Adamic and B. A. Huberman.
\newblock 2000.
\newblock The nature of markets in the World Wide Web. {\em Quarterly Journal of Electronic Commerce 1}, 512.

\bibitem[\protect\citename{Albert and Barab{\'a}si}2002]{Albert:02}
R. Albert and A.-L. Barab{\'a}si.
\newblock 2002.
\newblock Statistical mechanics of complex networks. {\em Reviews of Modern Physics 74}, 47--97. 

\bibitem[\protect\citename{Barab{\'a}si and Albert}1999]{Albert:99}
A.-L. Barab{\'a}si and R. Albert.
\newblock 1999.
\newblock Emergence of scaling in random networks. {\em Science 286}, 509-–512.

\bibitem[\protect\citename{de Boer}2000]{Boer:00}
Bart de Boer.
\newblock 2000.
\newblock Self-Organisation in Vowel Systems. {\em Journal of Phonetics}, Elsevier.

\bibitem[\protect\citename{Boersma}1998]{Boersma:98}
P. Boersma. 
\newblock 1998.
\newblock {\em Functional Phonology. (Doctoral thesis, University of Amsterdam)}, The Hague:
Holland Academic Graphics.

\bibitem[\protect\citename{Bulmer}1979]{Bulmer:79}
M. G. Bulmer.
\newblock 1979.
\newblock {\em Principles of Statistics}, Mathematics.  

\bibitem[\protect\citename{Cancho \bgroup et al.\egroup}2001]{Cancho:01}
Ferrer i Cancho and R. V. Sol{\'e}.
\newblock 2001.
\newblock Santa Fe working paper 01-03-016.

\bibitem[\protect\citename{Chomsky and Halle}1968]{Chomsky:68}
N. Chomsky and M. Halle.
\newblock 1968.
\newblock {\em The Sound Pattern of English}, New York: Harper and Row.
 
\bibitem[\protect\citename{Clements}2004]{Clements:04}
N. Clements.
\newblock 2004.
\newblock  Features and Sound Inventories. {\em Symposium on Phonological Theory: Representations and Architecture}, 
CUNY.

\bibitem[\protect\citename{Flemming}2002]{Flemming:02}
E. Flemming.
\newblock 2002.
\newblock {\em Auditory Representations in Phonology}, New York and London: Routledge.

\bibitem[\protect\citename{Gomes \bgroup et al.\egroup}1999]{Gomes:99}
M. A. F. Gomes, G. L. Vasconcelos, I. J. Tsang, and I. R. Tsang. 
\newblock 1999.
\newblock Scaling relations for diversity of languages. {Physica A}, 271, 489. 

\bibitem[\protect\citename{Greenberg}1966]{Greenberg:66}
J. H. Greenberg. 
\newblock 1966. 
\newblock {\em Language Universals with Special Reference to
Feature Hierarchies}, The Hague Mouton.

\bibitem[\protect\citename{Grossman \bgroup et al.\egroup}1995]{Grossman:95}
J. W. Grossman and P. D. F. Ion.
\newblock 1995. 
\newblock On a portion of the well-known collaboration graph. {\em Congressus Numerantium}, 108, 129–-131.

\bibitem[\protect\citename{Hinskens and Weijer}2003]{Hinskens:03}
F. Hinskens and J. Weijer.
\newblock 2003.
\newblock Patterns of segmental modification in consonant inventories: a cross-linguistic study. {\em Linguistics.}

\bibitem[\protect\citename{Jakobson}1941]{Jakobson:41}
R. Jakobson. 
\newblock 1941. 
\newblock {\em Kindersprache, Aphasie und allgemeine Lautgesetze}, Uppsala, Reprinted in {\em Selected Writings I.
Mouton}, The Hague, 1962, pages 328–-401.

\bibitem[\protect\citename{Jeong \bgroup et al.\egroup}2000]{Jeong:00}
H. Jeong, B. Tombor, R. Albert, Z. N. Oltvai, and A. L. Barab{\'a}si. 
\newblock 2000.
\newblock The large-scale organization of metabolic networks. {Nature}, 406:651–-654.

\bibitem[\protect\citename{Jakobson and Halle}1956]{Jakobson:56}
R. Jakobson and M. Halle. 
\newblock 1956.
\newblock {\em Fundamentals of Language}, The Hague: Mouton and Co.

\bibitem[\protect\citename{Ladefoged and Maddieson}1996]{Ladefoged:96}
P. Ladefoged and I. Maddieson.  
\newblock 1996.
\newblock {\em Sounds of the World's
Languages}, Oxford: Blackwell.

\bibitem[\protect\citename{Lindblom and Maddieson}1988]{Lindblom:88}
B. Lindblom and I. Maddieson. 
\newblock 1988.
\newblock Phonetic Universals in Consonant Systems. In L.M. Hyman and
C.N. Li, eds., {\em Language, Speech, and Mind}, Routledge, London, 62--78.

\bibitem[\protect\citename{Lotka}1926]{Lotka:26}
A. J. Lotka.
\newblock 1926.
\newblock The frequency distribution of scientific production. {\em J. Wash. Acad. Sci. 16}, 317–-323. 

\bibitem[\protect\citename{Maddieson}1984]{Maddieson:84}
I. Maddieson.
\newblock  1984.
\newblock {\em Patterns of Sounds}, Cambridge University Press, Cambridge.

\bibitem[\protect\citename{Martinet}1968]{Martinet:68}
A. Martinet.
\newblock 1968.
\newblock Phonetics and linguistic evolution. In Bertil Malmberg (ed.), {\em Manual of
phonetics, revised and extended edition}, Amsterdam: North-Holland Publishing Co. 464--487.

\bibitem[\protect\citename{Newman}2001b]{Newman:01}
M. E. J. Newman.
\newblock 2001b. 
\newblock Scientific collaboration networks. {\em I and II. Phys. Rev.}, E 64.

\bibitem[\protect\citename{Newman}2003]{Newman:03}
M. E. J. Newman.
\newblock 2003.
\newblock The structure and function of complex networks. {\em SIAM Review} 45, 167--256. 

\bibitem[\protect\citename{Pericliev and Vald{\'e}s-P{\'e}rez}2002]{Peri:02}
V. Pericliev, R. E. Vald{\'e}s-P{\'e}rez. 
\newblock 2002.
\newblock Differentiating 451 languages in terms of their segment inventories. {\em Studia Linguistica}, Blackwell 
Publishing.

\bibitem[\protect\citename{Pinker}1994]{Pinker:94}
S. Pinker.
\newblock 1994.
\newblock {\em The Language Instinct}, New York: Morrowo.

\bibitem[\protect\citename{Ramasco \bgroup et al.\egroup}2004]{Ramasco:04}
Jos{\'e} J. Ramasco, S. N. Dorogovtsev, and Romualdo Pastor-Satorras.
\newblock 2004.
\newblock Self-organization of collaboration networks. {\em Physical Review E}, 70, 036106.

\bibitem[\protect\citename{Simon}1955]{Simon:55}
H. A. Simon.
\newblock 1955.
\newblock On a class of skew distribution functions. {\em Biometrika 42}, 425–-440. 
 
\bibitem[\protect\citename{Trubetzkoy}1969/1939]{Trub:39}
N. Trubetzkoy. 
\newblock 1969.
\newblock {\em Principles of phonology. (English translation of Grundz{\"u}ge der
Phonologie, 1939)}, Berkeley: University of California Press.

\bibitem[\protect\citename{Vitevitch}2005]{Vitevitch:05}
M. S. Vitevitch. 
\newblock 2005.
\newblock Phonological neighbors in a small world: What can graph theory tell us about word learning? {\em Spring 2005 
Talk Series on Networks and Complex Systems}, Indiana University, Bloomington.

\bibitem[\protect\citename{Wang}1968]{Wang:68}
William S.-Y. Wang.
\newblock 1968.
\newblock The basis of speech, Project on Linguistic Analysis Reports, University of California at
Berkeley. Reprinted in {\em The Learning of Language}, ed. by C. E. Reed, 1971.

\bibitem[\protect\citename{Yook \bgroup et al.\egroup}2001b]{Yook:01}
S. Yook, H. Jeong and A.-L. Barab{\'a}si.
\newblock 2001b.
\newblock preprint.

\bibitem[\protect\citename{Zipf}1949]{Zipf:49}
G. K. Zipf.
\newblock 1949.
\newblock {\em Human Behaviour and the Principle of Least Effort}, Addison-Wesley, Reading, MA.

\end{thebibliography}
\end{document}